\RequirePackage{fix-cm}
\documentclass[smallextended]{svjour3}       
\smartqed  
\usepackage{graphicx}
 \usepackage{mathptmx}      
\usepackage{amsmath}
\newcommand{\m}{\phantom{-}}
\newcommand{\I}{{\rm i}}
\begin{document}

\title{The molecular nature of some exotic hadrons \thanks{G.M. and A.R.  acknowledge  support from the Spanish Ministerio de Econom\'ia y Competitividad (MINECO) under the project MDM-2014-0369 of ICCUB (Unidad de Excelencia ``Mar\'ia de Maeztu''), and, with additional European FEDER funds, under the contract FIS2017-87534-P. G.M. also acknowledges support from the FPU17/04910 Doctoral Grant from MINECO. 
A.F. acknowledges support from the GACR grant no. 19-19640S. Q.LL. acknowledges support from the Spanish Ministerio de Educaci\'on
y Ciencia, under the collaboration grant 18CO1/004638.}
}


\author{A. Ramos, 
        A. Feijoo, 
        Q. Llorens, 
        G. Monta\~na  
}


\institute{A. Ramos \at
              Departament de F\'{i}sica Qu\`{a}ntica i Astrof\'{i}sica and Institut de Ci\`{e}ncies del Cosmos, Universitat de Barcelona, Mart\'{i} i
Franqu\`{e}s 1, 08028 Barcelona, Spain \\
              \email{ramos@fqa.ub.edu}           
           \and
           A. Feijoo \at
              Nuclear Physics Institute, 25068 \v Re\v z, Czech Republic \\
              \email{feijoo@ujf.cas.cz}
              \and
     Q. Llorens \at
               Departament de F\'{i}sica Qu\`{a}ntica i Astrof\'{i}sica and Institut de Ci\`{e}ncies del Cosmos, Universitat de Barcelona, Mart\'{i} i
Franqu\`{e}s 1, 08028 Barcelona, Spain \\
\email{kimus7@gmail.com}
\and
   G. Monta\~na         \at
               Departament de F\'{i}sica Qu\`{a}ntica i Astrof\'{i}sica and Institut de Ci\`{e}ncies del Cosmos, Universitat de Barcelona, Mart\'{i} i
Franqu\`{e}s 1, 08028 Barcelona, Spain  \\
\email{gmontana@fqa.ub.edu}
}

\date{Received: date / Accepted: date}

\maketitle

\begin{abstract}
The exciting discovery by LHCb of the $P_c(4312)^+$ and  $P_c(4450)^+$ pentaquarks, or the suggestion of a tetraquark nature for the $Z_c(3900)$ state seen at BESIII and Belle, have triggered a lot of activity in the field of hadron physics, with new experiments planned for searching other exotic mesons and baryons, and many theoretical developments trying to disentangle the true multiquark nature from their possible molecular origin. After a brief review of the present status of these searches, this paper focusses on recently seen or yet to be discovered exotic heavy baryons that may emerge from a conveniently unitarized meson-baryon interaction model in coupled channels. In particular, we will show how interferences between the different coupled-channel amplitudes of the model may reveal the existence of a $N^*$ resonance around 2 GeV having a meson-baryon quasi-bound state nature. We also discuss the possible interpretation of
some of the $\Omega_c$ states recently discovered at LHCb as being hadron molecules. The model also predicts the existence of doubly-charmed quasibound meson-baryon $\Xi_{cc}$ states, which would be excited states of the ground-state  $\Xi_{cc}(3621)$ MeV, whose mass has only been recently established. Extensions of these results to the bottom sector will also be presented.

\end{abstract}

\section{Composite hadrons: a brief review}
\label{sec:historical}

The basic constituents of matter in QCD are quarks and gluons. Although hadrons, the colourless objects of the theory, could be realized in many complicated combinations of the elementary constituents, 
the conventional quark model proposed by Gell-Mann and Zweig \cite{GellMann:1964nj,Zweig:1964}, according to which mesons are composed by a quark-antiquark pair and baryons are built from a three-quark cluster, has successfully described the phenomenological observations of ground state hadrons
\cite{Godfrey:1985xj,Capstick:1986bm}. Nonetheless, a great deal of theoretical and experimental activities in hadron physics have been focussed on finding evidence of exotic components in the mesonic and baryonic spectrum for more than fifty years \cite{Klempt:2007cp,Crede:2008vw,Brambilla:2014jmp}.


There exist many theoretical predictions and experimental candidates of exotic mesons \cite{Amsler:2004ps}. These include glueballs (objects made entirely of gluons), tetraquarks (compact systems of two quarks and antiquarks) or two-meson molecules bound by an attractive interaction. The constituent quark model fails already at explaining the large mass difference of some low lying states, such as the flavour-equivalent $L=1$ spatial excited mesons, $f_0(500)$ and $a_0( 980)$.  A natural explanation is provided by unitarized theories of pseudoscalar meson scattering employing chiral lagrangians \cite{Oller:1997ti,Oller:1998hw,Pelaez:2015qba}, from which the $f_0(500)$ emerges as a $\pi\pi$ resonance, while the $a_0(980)$ and its $f_0(980)$ partner correspond mainly to quasi-bound states of $K{\bar K}$ pairs in isospin $I=1$ and $I=0$, respectively.  

One of the early evidences for a composite or molecular type baryon was provided by the $\Lambda(1405)$ resonance, whose mass was systematically predicted  to be too high by quark models but it found a better explanation if it was described as a $\bar K N$ quasi-bound state. This was already pointed out in the late fifties \cite{Dalitz:1960du}, and was later corroborated by models that built the meson-baryon interaction from a chiral effective lagrangian and  implemented unitarization \cite{Kaiser:1995eg,Oset:1997it,Oller:2000fj,Ikeda:2012au,Feijoo:2018den}. A clear indication that the $\Lambda(1405)$ is a meson-baryon quasibound state comes from the confirmation of its related double-pole nature \cite{Oller:2000fj,Jido:2003cb} after comparing different experimental line shapes \cite{Magas:2005vu} having selectivity to either one or the other pole.

Disentangling the true nature of a particular hadron is not easy
due to the mixing of conventional and exotic components. The situation has changed
since the beginning of the millenium,  as higher energies became available in the experimental facilities giving access to the heavy flavored meson and baryon resonance region \cite{Zyla:2020}. Many of the new XYZ meson resonances produced could not be explained as having the $q{\bar q}$ structure of the conventional quark model \cite{Godfrey:2008nc}, being notable the {\it charged} charmonium states, $Z^{\pm}_c$ and $Z^{\pm}_b$ \cite{Ablikim:2013mio} which definitely need the presence of an additional $q{\bar q}$ pair. 
Of particular importance is the $X(3872)$ meson \cite{Choi:2003ue} , as it was
the first charmonium state that did not fit a $c \bar c$ picture and stimulated the search of alternative explanations.
Some models advocate for a molecular
$D D^{*}$ nature \cite{mol1,mol2}, due to the closeness of the $X(3872)$ mass to the $D D^{*}$ threshold, while other models describe it as having a bound diquark and antidiquark (tetraquark) structure \cite{tetra1,tetra2}. 
Similarly, in the baryonic sector, the excited nucleon resonances  $P_c(4312)$ and  $P_c(4450)$, recently observed at LHCb \cite{Aaij:2015tga,Aaij:2019vzc} in the invariant mass distribution of $J/\psi \, p$ pairs from the decay of the $\Lambda_b$, find a natural explanation in terms of a pentaquark structure. Some models had already predicted the existence of highly excited nucleonic states as being bound systems generated from the dynamics of the $J/\psi N$ interaction and related coupled channels \cite{Wu:2010jy,Yuan:2012wz,Xiao:2013yca,Garcia-Recio:2013gaa}, and they have been recently shown to accommodate the LHCb observations \cite{Roca:2015dva}. Suggestions for finding the $S=-1$ partners already predicted in \cite{Wu:2010jy,Yuan:2012wz},  through the observation of appropriate meson-baryon pairs produced in the decay of bottom baryons at LHCb, have also been given \cite{Chen:2015sxa,Feijoo:2015kts}.

For an overview of the situation and a detailed discussion of the various models see the reviews~\cite{Brambilla:2010cs,Esposito:2014rxa,Guo:2017jvc} and references therein.

After reviewing the formalism of the model employed in this work, we focus on recent experimental reaction cross sections, in the light and in the charm sectors, that could be interpreted through the presence of baryon resonances made of five quarks in the form of meson-baryon quasibound states. We will also give predictions for possible meson-baryon excited resonances in the doubly charmed baryonic sector.

\section{Formalism}
\label{sec:formalism}

\begin{figure}[h!]
\centering
 \includegraphics[width=0.80\textwidth]{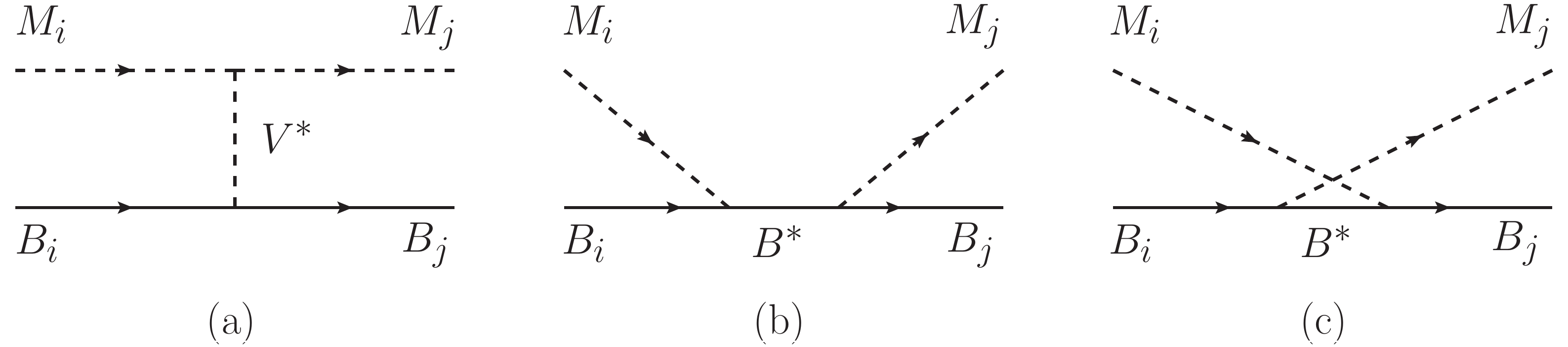}
\caption{Leading order tree level diagrams contributing to the $MB$ interaction. Baryons and mesons are depicted by solid and dashed lines, respectively.}
\label{fig-t}
\end{figure}

The meson-baryon interaction model employed in this work is based on the tree-level diagrams of Fig.~\ref{fig-t}. The s-wave interaction kernel $V_{ij}$ is obtained from the t-channel vector meson exchange amplitude of diagram (a) \cite{Hofmann:2005sw}:
\begin{equation}\label{eq:Vij}
 V_{ij}(\sqrt{s})=-C_{ij}\frac{1}{4f^2}\left(2\sqrt{s}-M_i-M_j\right) N_i N_j\, ,
\end{equation}
with $M_i$, $M_j$ and $E_i$, $E_j$ being the masses and the energies of the baryons, and $N_i$, $N_j$ the normalization factors ${N=\sqrt{(E+M)/2M}}$.
The coefficients $C_{ij}$ are obtained from the evaluation of the diagram, in the $t\ll m_V$ limit, employing effective Lagrangians of the hidden gauge formalism:
\begin{equation}\label{eq:vertexVPP}
\mathcal{L}_{VPP}=ig\langle\left[\partial_\mu\phi, \phi\right] V^\mu\rangle\,,
\end{equation}
\begin{equation}\label{eq:vertexBBV}
\mathcal{L}_{VBB}=\frac{g}{2}\sum_{i,j,k,l=1}^4\bar{B}_{ijk}\gamma^\mu\left(V_{\mu,l}^{k}B^{ijl}+2V_{\mu,l}^{j}B^{ilk}\right)\, ,
\end{equation}
that describe the vertices coupling the vector meson to pseudoscalars ($VPP$) and baryons ($VBB$), respectively, in the scattering of pseudoscalar mesons off baryons ($PB$). The coupling constant $g$ is related to the pion decay constant $f$ and a representative vector meson mass $m_V$, taken as the $\rho$ meson mass, by $g=m_V/2f$.

The interaction of vector mesons with baryons ($VB$) is built in a similar way and involves the three-vector $VVV$ vertex, which is obtained from:
\begin{equation}\label{eq:vertexVVV}
\mathcal{L}_{VVV}=ig\langle {\left[V^\mu,\partial_\nu V_\mu\right] V^\nu}\rangle \,.
\end{equation}
The resulting $VB$ interaction is that of Eq.~(\ref{eq:Vij}) with the addition of the product of polarization vectors, $\vec{\epsilon}_i\cdot\vec{\epsilon}_j$.

In the above expressions, $\Phi$, $V$ and $B$ represent the pseudoscalar, vector, and baryon tensor fields built up from $u,d,s$ and $c$ quarks (labelled $i=1,2,3,4$, respectively). 
Despite $SU(4)$ symmetry is incorporated in the Lagrangians, the use of the physical hadron masses leads to an explicit SU(4) symmetry breaking in the corresponding kernel. Further, a factor $\kappa_c \sim1/4$ is implemented in the non-diagonal transitions mediated by the exchange of a charm vector meson to account for its higher mass with respect to the light ones.  We note that the transitions in which a light vector meson is exchanged, like the dominant diagonal ones, do not make explicit use of $SU(4)$ symmetry since they are effectively projected into their $SU(3)$ content.

The s-channel and u-channel diagrams of Figs.~\ref{fig-t}(b) and (c), usually referred to as Born terms, have been widely studied in $S=-1$ sector. Given the chiral nature of the vertices in that sector, these terms contribute mainly in p-wave and their s-wave contribution is almost negligible at energies around the lowest threshold. Nevertheless, as pointed out in Ref.~\cite{Oller:2000fj}, they may reach 20\% of the leading order (LO) contact term of Eq.~\ref{eq:Vij} at energies 200~MeV above threshold. A clear proof of the relevant role played by the s- and u-channels in s-wave at higher energies can be seen in Ref.~\cite{Ramos:2016odk}. This phenomenology might naively skew one's criteria to justify omitting the Born terms from the LO calculations in other sectors.  However, one should be cautious and not only pay attention on the energy range but also on the symmetry-based structures required by the sector in the vertices of the Born diagrams \cite{albert}, as will be illustrated in one of the examples discussed in this work.  

The sought resonances are dynamically generated as poles of the scattering amplitude $T_{ij}$, unitarized by means of the on-shell Bethe-Salpeter equation in coupled channels, which implements the resummation of loop diagrams to infinite order:
\begin{equation}\label{eq:BSeq}
T_{ij}=V_{ij} + V_{il}G_lV_{lj} + V_{il}G_lV_{lk}G_kV_{kj} + \cdots = V_{ij}+V_{il}G_{l}T_{lj}\, .
\end{equation}
In the on-shell factorization approach, the former integral equation becomes an algrebraic one and the meson--baryon loop function $G_{l}$ is given by:
\begin{equation}
    G^I_l = i \int \frac{d^4q}{(2\pi)^4}\frac{2M_l}{(P - q)^2 - M_l^2 + i\epsilon}\frac{1}{q^2 - m_l^2 + i\epsilon}\,,
\end{equation}
where $M_l$ and $m_l$ are, respectively, the masses of the baryon and meson in the loop. The ultraviolet divergence of the loop
is regularized using the \textit{dimensional regularization} approach, which introduces a subtraction constant $a_l(\mu)$ for each intermediate channel $l$ at a given regularization scale $\mu$. Alternatively, one could also employ a cut-off regularization scheme. For a proper physics interpretation of the results, it is convenient to demand ``natural" values of the subtraction constants \cite{Oller:2000fj}, obtained by matching the loop function calculated by means of dimensional regularization with the one employing a cut-off whose value should remain of the order of the $\rho$-meson mass, as this is the scale of the degrees of freedom integrated out when reducing the t-channel vector-meson exchange diagram to a contact term.


The poles of the scattering amplitude $T_{ij}$ are to be found in the second Riemann sheet of the complex plane, defined 
as the one which is connected to the real energy axis from below and obtained by employing:
\begin{equation}
G^l_{\rm II}(\sqrt{s}+i\varepsilon)=G^l_{\rm I}(\sqrt{s}+i\varepsilon)+i\frac{q_l}{4\pi\sqrt{2}} \ ,
\label{2ndRiemann}
\end{equation}
for ${\rm Re } \sqrt{s}> m_{l} + M_{l}$ and $G^l_{\rm I}$ for ${\rm Re } \sqrt{s}< m_{l} + M_{l}$.

Around the pole position, the scattering amplitude can be approximated by the expression
\begin{equation}
T^{ij}(z)=\frac{g_ig_j}{z-z_p}\, ,
\end{equation}
from which one can derive the coupling constants ($g_i$) of the resonance or bound state to the channel $i$. In addition, following the idea of compositeness of shallow bound states formulated for the deuteron in Ref.~\cite{Weinberg:1962hj}, and extending it to resonances with open channels for decay \cite{Aceti:2014ala}, one may assign the quantity 
\begin{equation}
\chi_i= \left|g_i^2\frac{\partial G_i(z_p)}{\partial z}\right| \ ,
\label{eq:compo}
\end{equation}
to the amount of $i^{\rm th}$ channel meson-meson component contained in the the dynamically generated state \cite{Guo:2015daa}.


\section{A selected new case in SU(3)}

The interpretation of recent data of photoproduction reactions in a center-of-mass energy region of around 2 GeV has led to exploring the relevance of vector mesons in unitarized coupled channel models for the dynamical generation of resonances. A nucleon resonance around 1970~MeV and width around 65 MeV, coupling mostly to $K^*\Lambda$ and $K^* \Sigma$ states, was first reported in \cite{angelsvec}.  The work of
\cite{Ramos:2013wua} demonstrated that this resonance, together with the coupled channel vector-baryon dynamics, provided an explanation of the features observed by the CBELSA/TAPS collaboration in the $\gamma p \to K^0 \Sigma^+$ cross section \cite{schmieden} , namely a peak around 
$\sqrt s =1900$~MeV followed by a fast downfall around $\sqrt s =2000$~MeV.

\begin{figure}[ht]
\begin{minipage}[t]{0.47\linewidth}
\centering
\includegraphics[width=0.63\textwidth]{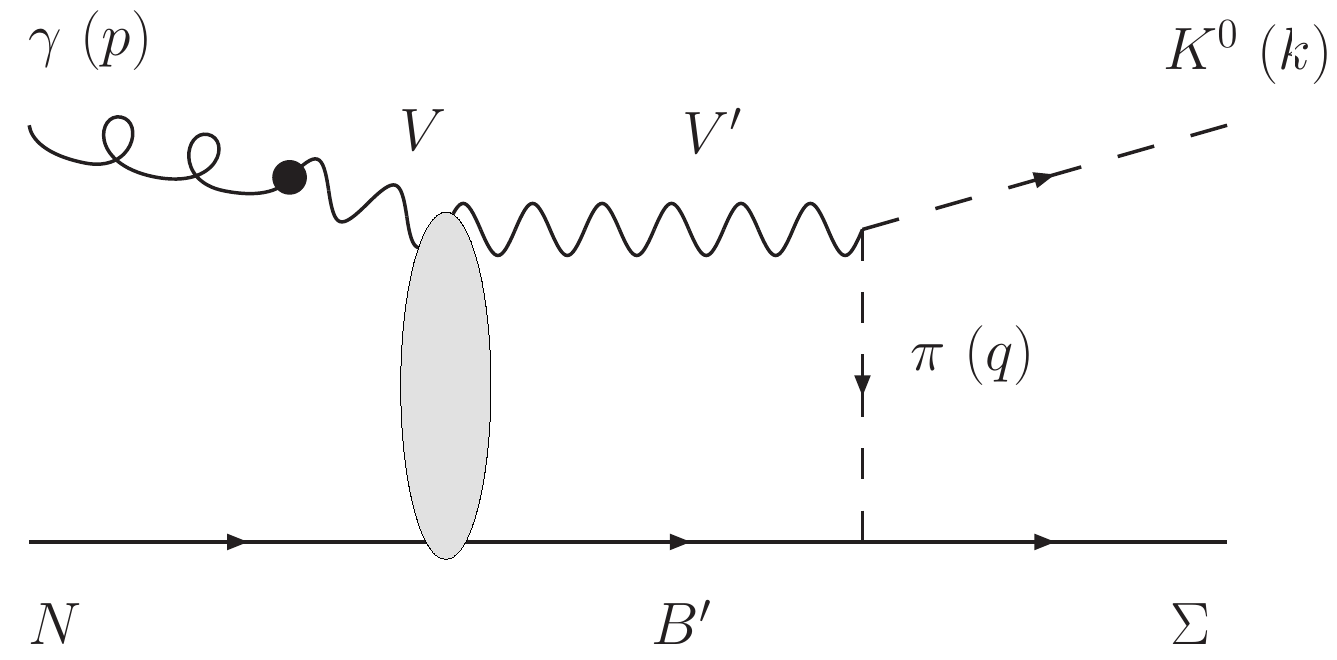}
\caption{Mechanism for the photoproduction reaction $\gamma N \to K^0\Sigma$.}
\label{fig:diag}
\end{minipage}
\hspace{0.5cm}
\begin{minipage}[t]{0.47\linewidth}
\centering
\includegraphics[width=0.63\textwidth]{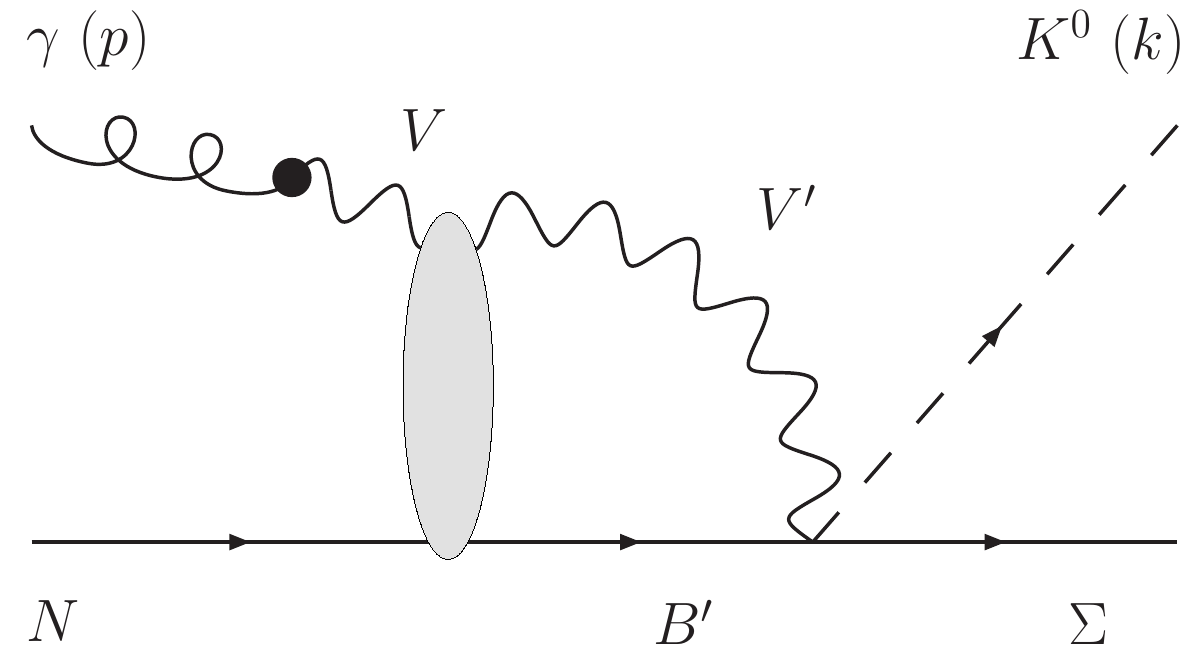}
\caption{Kroll-Ruderman contact term, added to the mechanisms of
Fig.~\protect\ref{fig:diag} to preserve gauge-invariance. }
\label{fig:KR}
\end{minipage}
\end{figure}

The role of coupled-channel unitarization is particularly significant in this case because the tree-level t-channel term is zero.
The basic mechanisms of the model of Ref.~\cite{Ramos:2013wua} are depicted in Figs.~\ref{fig:diag} and \ref{fig:KR}, where one can see the
photon conversion into a neutral vector meson ($\rho^0, \omega, \phi$) followed by the $\rho N, \omega N,
\phi N$
interaction leading to the relevant vector-baryon ($V^\prime B^\prime$)
channels, which are restricted to be $K^*\Lambda$ or  $K^* \Sigma$,
since those are
the ones to which the resonance around 1970~MeV couples most strongly. The vector-baryon unitarized amplitudes are taken from the work
of Ref.~\cite{angelsvec}, which builds the required interaction Lagrangians from the hidden gauge formalism.
Finally,  the
intermediate $V^\prime B^\prime$ states get converted via pion
exchange to the $K^{0} \Sigma$ final state.  

\begin{figure}[ht]
\begin{minipage}[t]{0.47\linewidth}
\centering
\includegraphics[width=0.95\textwidth]{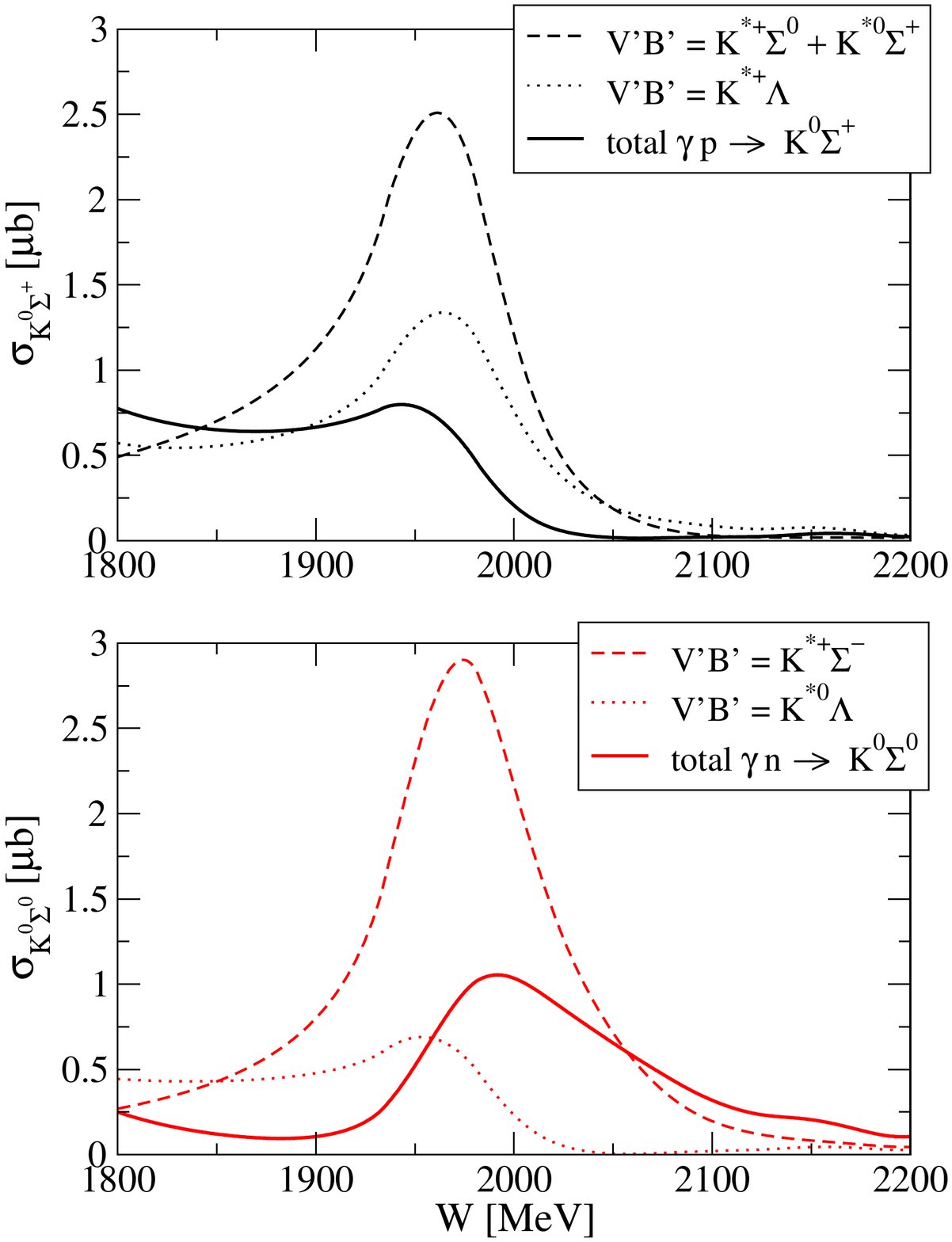}
\caption{Contributions to the $\gamma p \to K^0 \Sigma^+$ (upper panel) and 
$\gamma n \to K^0 \Sigma^0$ (lower panel) cross sections. }
\label{fig:gn_gp}
\end{minipage}
\hspace{0.5cm}
\begin{minipage}[t]{0.47\linewidth}
\centering
\includegraphics[width=0.95\textwidth]{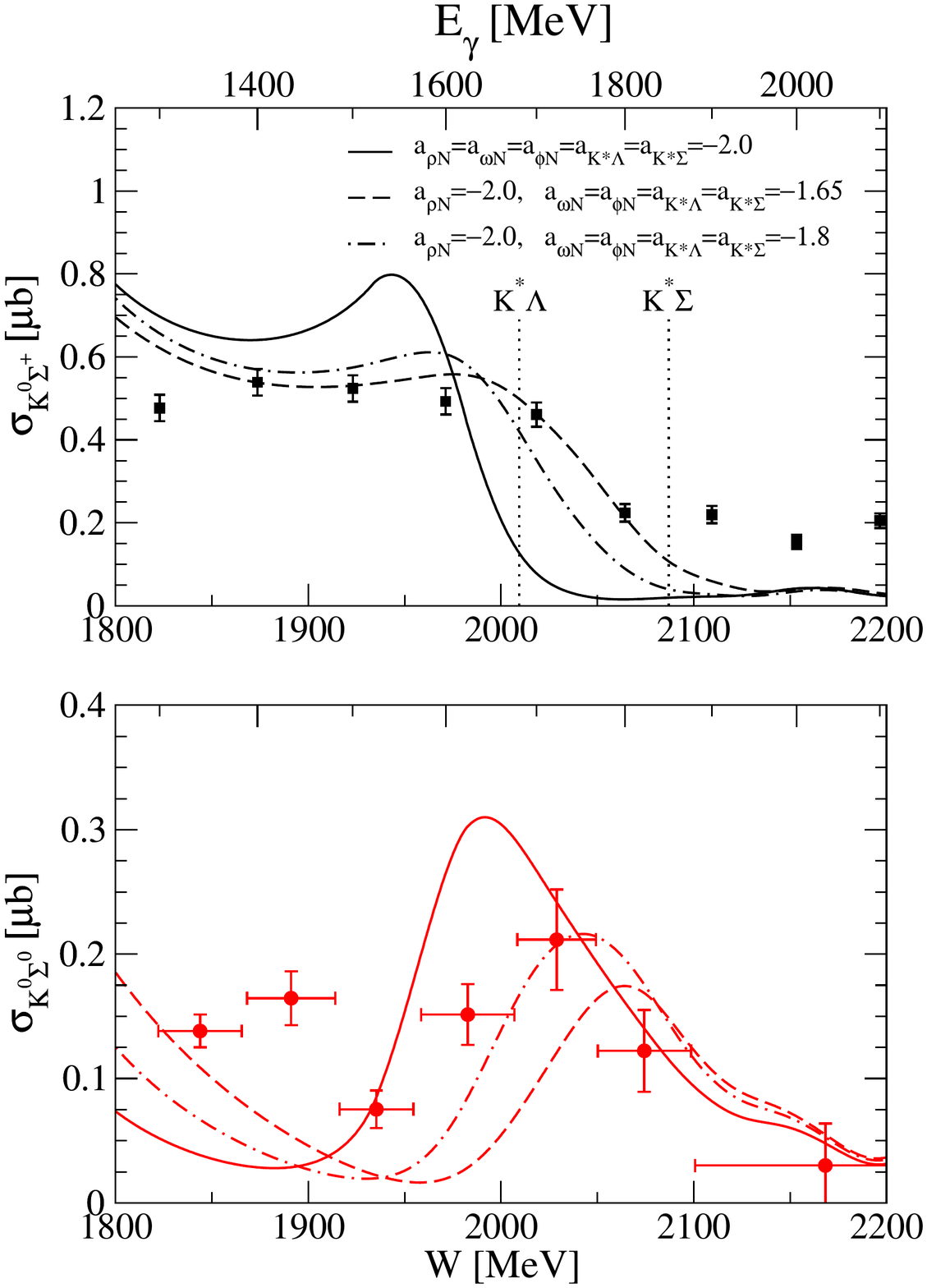}
\caption{Comparison of the $\gamma p \to K^0 \Sigma^+$ cross
section, obtained with two parameter sets, with the CBELSA/TAPS data of
Ref.~\protect\cite{schmieden} (upper panel).
Scaled predictions for the $\gamma n \to K^0 \Sigma^0$ cross section using
two parameter sets, compared with the recent BGO-OD data \cite{Jude:2020bwn,Kohl:BGOOD} (lower panel). }
\label{fig:sigma}
\end{minipage}
\end{figure}

The results shown in Fig.~\ref{fig:gn_gp}, obtained by retaining different intermediate channels, demonstrate that there is a destructive interference between the $V B \to K^{*}\Sigma,K^*
\Lambda$ amplitudes, 
which are of similar
 size and  shape in the case of  the $\gamma p \to K^0 \Sigma^+$ reaction. 
This
produces an
abrupt downfall of the cross section, right at the position of the resonance
generated by the employed $VB$ interaction model. In contrast, the
$\gamma n \to K^0 \Sigma^0$ cross section retains the peak at the position of
the resonance. 
The downfall of the $\gamma p \to K^0 \Sigma^+$ cross section
(solid line)  appears 60~MeV below the
energy at which the experimental cross section presents the abrupt drop, 
as can be seen in
the upper panel of Fig.~\ref{fig:sigma}. 
Since this structure is sensitive to the position of the resonance, a prediction of its properties can be obtained by adjusting the parameters of the model to reproduce the CBELSA/TAPS data. The fitted result, shown by a dashed line in Fig.~\ref{fig:sigma}, moves the resonance up to 2030~MeV, which then lays above the $K^* \Lambda$ threshold and becomes almost twice wider than in the original model. 
These results are backed up by two resonances of negative parity around this region of energy,
$N^*(2080) (3/2^-)$ and $N^*(2090)(1/2^-)$, that appeared in earlier
versions of the PDG, and by a $3/2^-$ state around
2080 MeV found to explain SPring 8 LEPS data on the
$\gamma p \to K^+ \Lambda(1520)$ reaction in \cite{Xie:2010yk}. 
This is further supported by the recent
measurement of the neutral $\gamma n\to K^0 \Sigma^0$ cross section by the BGO-OD collaboration \cite{Jude:2020bwn,Kohl:BGOOD}, shown in the 
the lower panel of Fig.~\ref{fig:sigma}, together with the prediction of the model for
several parameter sets. The theoretical results have been conveniently scaled to match the data as the later are limited to a particular angular bin, $0.2 < \cos\theta^K_{\rm c.m.}  < 0.5$.  
The experimental enhancement around  2050 MeV is consistent with the structure 
that remains in the theoretical cross section of the $\gamma n \to K^0 \Sigma^0$ reaction, since in this case the 
interference of amplitudes is dominated by that involving intermediate $K^*\Sigma$ channels that shows a clear resonant peak. Note that the parameter set that produces the the dot-dashed line, adjusting better to the BGO-OD data, gives a poorer but yet acceptable description of the charged $K^0 \Sigma^+$ channel.

\section{Baryon resonances with charm}

\subsection{The $(I,S,C)=(0,-2,1)$ sector: $\Omega^0_c$}
\label{sec:omega}

The discovery of five narrow $\Omega_c^0$ excited resonances by the LHCb Collaboration \cite{Aaij:2017nav} has increased the 
theoretical effort in the field of baryon spectroscopy in the charm sector with the aim of explaining their inner structure 
and possibly establishing their unknown values of spin--parity. While some works suggest a $css$ quark description within revisited quark models \cite{Agaev:2017jyt,Chen:2017sci,Karliner:2017kfm,Wang:2017hej,Wang:2017vnc,Cheng:2017ove,Wang:2017zjw,Chen:2017gnu,Santopinto:2018ljf},
others propose a pentaquark interpretation \cite{Huang:2017dwn,An:2017lwg,Kim:2017jpx}. Earlier
models that obtain $\Omega_c^0$ resonances as quasi-bound states of an interacting meson--baryon pair \cite{Hofmann:2005sw,JimenezTejero:2009vq,Romanets:2012hm} have been recently re-examined \cite{Montana:2017kjw,Debastiani:2017ewu,Wang:2017smo,Chen:2017xat,Nieves:2017jjx} in view of the new experimental data. 
This possibility is supported by the fact that the masses of the excited $\Omega_c^0$ baryons under study lie near the $\bar K\Xi_c$ and $\bar K\Xi_c^\prime$ thresholds and that they have been observed in the $K^-\Xi^+_c$ invariant mass spectrum.

The available $PB$ channels in the $(I,S,C)=(0,-2,1)$ sector are $\bar{K}\Xi_c (2965)$, $\bar{K}\Xi'_c (3072)$, $D\Xi (3185)$, $\eta \Omega_c (3245)$, $\eta' \Omega_c (3655)$, $\bar{D}_s \Omega_{cc} (5519)$, and $\eta_c \Omega_c (5677)$, with the corresponding threshold masses in parenthesis. The doubly charmed $\bar{D}_s \Omega_{cc}$ and $\eta_c \Omega_c $ channels are neglected as their energy is much larger than that of the other channels. 
The matrix of $C_{ij}$ coefficients for the resulting 5-channel interaction is given in Table~\ref{tab:coeff}. Similarly, in the $VB$ case, the allowed states are $D^*\Xi (3326)$, $\bar{K}^*\Xi_c (3363)$, $\bar{K}^*\Xi'_c (3470)$, $\omega \Omega_c (3480)$, $\phi \Omega_c (3717)$, $\bar{D}_s^* \Omega_{cc} (5662)$ and $J/\psi \Omega_c (5794)$, where, again, the doubly charmed states are neglected. The corresponding  $C_{ij}$ coefficients can be straightforwardly obtained from those in Table~\ref{tab:coeff} with: $\pi \rightarrow \rho,\, K\rightarrow K^\ast,\, \bar{K}\rightarrow\bar{K}^*,\,  D\rightarrow D^*,\, \bar{D}\rightarrow\bar{D}^*,\, {1/\sqrt{3}}\eta+\sqrt{2/3}\eta'\rightarrow\omega$ and $-\sqrt{2/3}\eta+{1/\sqrt{3}}\eta'\rightarrow\phi$.

\begin{table}[h]
\caption{The $C_{ij}$ coefficients for the $(I,S,C)=(0,-2,1)$ sector of the  $PB$ interaction.}
\begin{center}
\begin{tabular}{l c c c c c}
\hline \\   [-3mm]
&{${\bar K}\Xi_c$}  & {${\bar K}\Xi_c^\prime$}  & { $D\Xi$}  & { $\eta\Omega_c^0$} &{$\eta^\prime\Omega_c^0$}  \\
\hline \\   [-3mm]
{${\bar K}\Xi_c$}         & $1$ & $0$ & $\sqrt{3/2}~\kappa_c$  & $\m\m 0$ & $\m\m 0$     \\
{${\bar K}\Xi_c^\prime$}   &     & $1$ & $\sqrt{1/2}~\kappa_c$ & $-\sqrt{6}$ & $\m\m 0$   \\
{ $D\Xi$} &     &     &    $\m 2$        &  $-\sqrt{1/3}~\kappa_c$  & $-\sqrt{2/3}~\kappa_c$ \\
{ $\eta\Omega_c^0$} &     &     &     &  $\m\m 0$  &  $\m\m 0$\\
{ $\eta^\prime\Omega_c^0$}  &     &     &    &     &  $\m\m 0$   \\             
\hline \\
\end{tabular}
\end{center}
\label{tab:coeff}
\vspace{-1.0cm}
\end{table}

Choosing the values of the subtraction constants, $a_l(\mu=1\rm~GeV)$, so that the loop function in dimensional regularization matches the one regularized with a cut-off $\Lambda=800\rm\,MeV$, gives rise to two poles in the $PB$ scattering amplitude, with energies that are similar to the second and fourth $\Omega_c^0$ states discovered by LHCb \cite{Aaij:2017nav}. However,
as the mass of the obtained heavier state is larger by 10\,MeV and its width is about twice the experimental values, we let the five subtraction constants to vary freely within a reasonably constrained range and look for a combination that reproduces the characteristics of the two observed states, $\Omega_c^0(3050)$  ($\Gamma=0.8\pm0.3$)
and $\Omega_c^0(3090)$  ($\Gamma=8.7\pm1.8$), within $2\sigma$ of the experimental errors. 
Table~\ref{tab:pseudo} displays the properties of the poles for a representative set of $a_l(\mu=1\rm~GeV)$ with equivalent cut-off values in the $320$--$950\;\rm MeV$ range, corresponding to ``Model 2" in ref.~\cite{Montana:2017kjw}. We note that the strongest change corresponds to $a_{\bar{K}\Xi_c}$, needed to decrease the width of the $\Omega_c^0(3090)$. Its equivalent cut-off value of 320 MeV is on the low side of the usually employed values but  can still be considered naturally sized. The results of Table~\ref{tab:pseudo} show that the main component of the lowest energy state at 3050\,MeV is $\bar{K}\Xi'_c$, although it also couples appreciably to $D\Xi $ and $\eta \Omega_c$ states. The higher energy resonance at 3090\,MeV couples strongly to $D\Xi$ and clearly qualifies as a $D\Xi $ bound state with a corresponding compositeness of 0.91.

\begin{table}[hbt!]
\caption{Position ($\sqrt{s}=M-\I\Gamma/2$), couplings and compositeness of the $\Omega^0_c$ states.}
\begin{center}
\begin{tabular}{lccccc}
\hline 
& &   \multicolumn{2}{c}{}    &   \multicolumn{2}{c}{} \\   [-3mm]
\multicolumn{6}{c}{ {\bf $0^- \otimes 1/2^+$} interaction in the {\bf$(I,S,C)=(0,-2,1)$} sector } \\
\hline
\multicolumn{6}{c}{ diagram (a) (Model 2 of \cite{Montana:2017kjw},) } \\
\hline
&  &  \multicolumn{2}{c}{}    &   \multicolumn{2}{c}{} \\   [-3mm]
$M\;\rm[MeV]$       &      &    \multicolumn{2}{c}{$\qquad 3050.3$}    &    \multicolumn{2}{c}{$\qquad 3090.8$}   \\
$\Gamma\;\rm[MeV]$  &  &     \multicolumn{2}{c}{$\qquad 0.44$}        &    \multicolumn{2}{c}{$\qquad 12$}     \\ 
\hline
 &  &  \multicolumn{2}{c}{}    &   \multicolumn{2}{c}{} \\  [-3mm]
  &$a_l $  &   $\qquad | g_i|$    & $\chi_i$    &   $\qquad | g_i|$  & $\chi_i$ \\ 
$\bar{K}\Xi_c (2965)$ &  $-1.69$   &  $\qquad 0.11$  & $0.00$     &    $\qquad 0.49$  & $0.02$  \\
$\bar{K}\Xi'_c (3072)$ &  $-2.09$ &  $\qquad 1.80$  & $0.61$     &    $\qquad 0.35$  & $0.03$  \\
$D\Xi (3185)$  &  $-1.93$         &  $\qquad 1.36$  & $0.07$     &    $\qquad 4.28$  & $0.91$ \\
$\eta \Omega_c (3245)$ &  $-2.46$ &  $\qquad 1.63$  & $0.14$     &    $\qquad 0.39$  & $0.01$ \\
$\eta' \Omega_c (3655)$&  $-2.42$ &  $\qquad 0.06$  & $0.00$     &    $\qquad 0.28$  & $0.00$ \\
\hline
\end{tabular}
\begin{tabular}{lccccccc}
\multicolumn{8}{c}{diagrams (a)+(b)+(c) ( \cite{albert} ) } \\
\hline
&  &  \multicolumn{2}{c}{} &  \multicolumn{2}{c}{}    &   \multicolumn{2}{c}{} \\   [-3mm]
$M\;\rm[MeV]$       &   &  \multicolumn{2}{c}{$\qquad 2997.4$}   &    \multicolumn{2}{c}{$\qquad 3045.5$}    &    \multicolumn{2}{c}{$\qquad 3070.1$}   \\
$\Gamma\;\rm[MeV]$  &  &     \multicolumn{2}{c}{$\qquad 14.4$}        &    \multicolumn{2}{c}{$\qquad 1.7$}  &     \multicolumn{2}{c}{$\qquad 8.4$}      \\ 
\hline
 &  &  \multicolumn{2}{c}{} &  \multicolumn{2}{c}{}    &   \multicolumn{2}{c}{} \\  [-3mm]
  &$a_l$  &   $\qquad | g_i|$    & $\chi_i$     &   $\qquad | g_i|$    & $\chi_i$    &   $\qquad | g_i|$  & $\chi_i$ \\   
$\bar{K}\Xi_c (2965)$ &  $-3.54$   &  $\qquad 0.80$  & $0.11$     &    $\qquad 0.04$  & $0.00$  &    $\qquad 0.12$  & $0.00$ \\
$\bar{K}\Xi'_c (3072)$ &  $-1.71$ &  $\qquad 0.04$  & $0.00$     &    $\qquad 1.05$  & $0.18$  &    $\qquad 1.36$  & $0.84$  \\
$D\Xi (3185)$  &  $-1.75$         &  $\qquad 0.06$  & $0.00$     &    $\qquad 4.60$  & $0.82$   &    $\qquad 4.57$  & $0.91$\\
$\eta \Omega_c (3245)$ &  $-2.31$ &  $\qquad 0.09$  & $0.00$     &    $\qquad 0.61$  & $0.02$  &    $\qquad 1.75$  & $0.17$\\
$\eta' \Omega_c (3655)$&  $-1.80$ &  $\qquad 0.08$  & $0.00$     &    $\qquad 0.43$  & $0.00$  &    $\qquad 0.30$  & $0.00$\\
\hline
\end{tabular}
\end{center}
\label{tab:pseudo}
\vspace{-0.7cm}
\end{table}

\begin{figure}[h]
\vspace{0.5cm}
\includegraphics[width=0.57\textwidth]{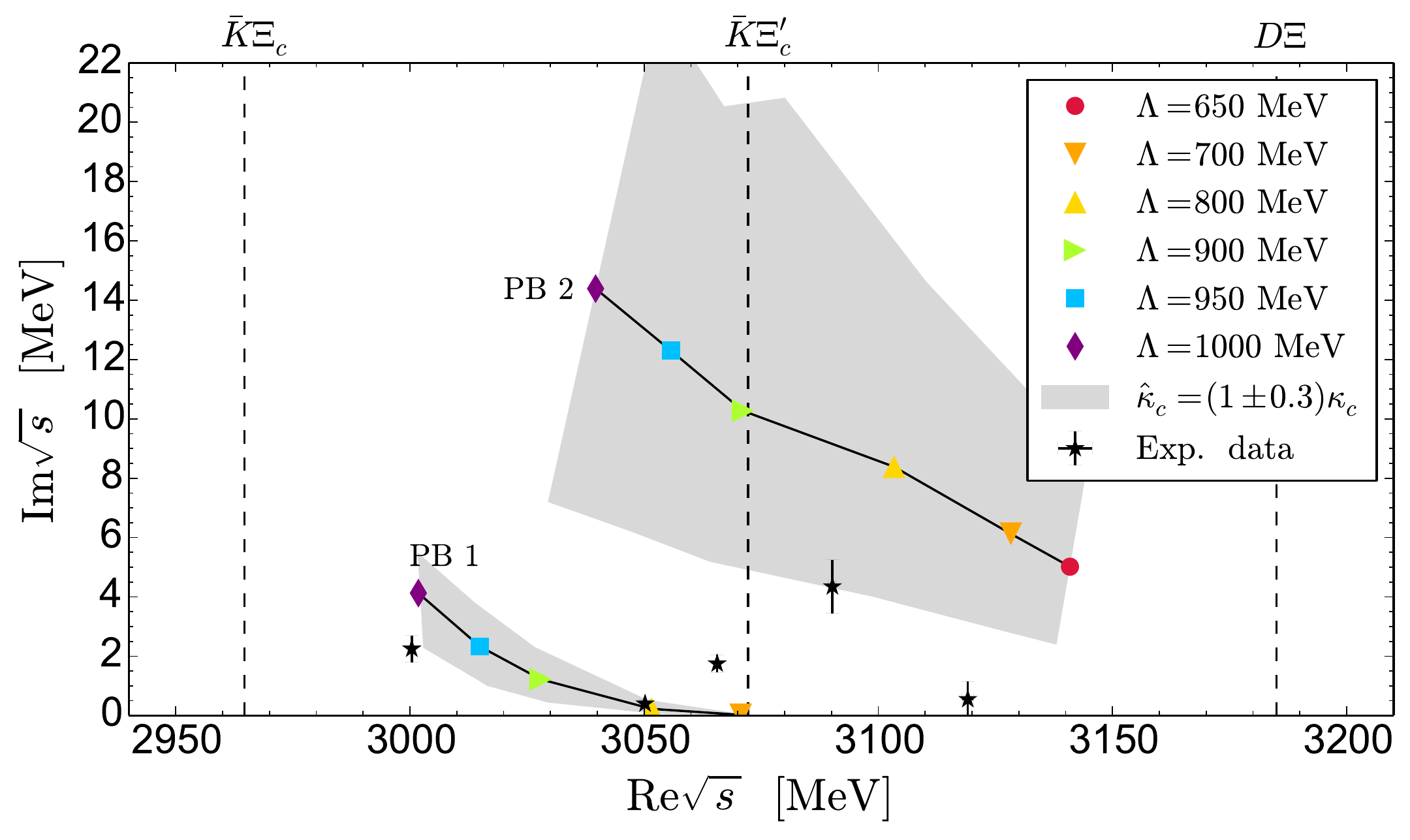}\hspace{0.5cm}
\begin{minipage}[b]{0.38\textwidth}\vspace{0.2cm} \caption{Evolution of the position of the resonance poles for various cut-off values. The grey area indicates the region of results covered when a variation of 30\% in the $SU(4)$ breaking is assumed in the transitions mediated by heavy-meson exchange.}\vspace{0.5cm}\label{fig:cut-off}
\end{minipage}
\vspace{-0.5cm}
\end{figure}

We performed a more systematic study of the 
dependence of these results on the assumed value of the cut-off, as well as the influence of a certain amount of $SU(4)$ symmetry violation associated to the fact that the charm quark is substantially heavier than the light quarks. 
The solid lines in Fig.~\ref{fig:cut-off} indicate the evolution of the poles as the value of the cut-off is increased from 650\,MeV to 1000\,MeV. An additional violation of $SU(4)$ symmetry, which is already broken by the use of the physical meson and baryon masses in the interaction kernel, is enforced by allowing a variation of up to $30\%$  in the coefficients of transitions mediated by the exchange of a charmed vector meson. The grey area in Fig.~\ref{fig:cut-off} displays the region in the complex plane where the resonances can be found varying both the cut-off and the amount of $SU(4)$ violation. The fact that these bands of uncertainties include the 3050\,MeV and the 3090\,MeV resonances measured at LHCb reinforces their interpretation as meson--baryon molecules.

We have taken a step forward in the interaction model by including the s-wave contribution of the usually neglected s-channel and u-channel Born terms (diagrams (b) and (c) in Fig.~\ref{fig-t}). While the details of the methodology and a systematic analysis of the results will be presented elsewhere \cite{albert}, we anticipate in the present work some interesting preliminary results that can be seen in the bottom part of Table~\ref{tab:pseudo}. Contrary to naive expectations, it is found that the Born terms in s-wave have a significant impact in the unitarized amplitudes. 
Instead of the two poles of the earlier model \cite{Montana:2017kjw}, we now find three poles that can possibly be identified, by its energy position and width, with the first three  $\Omega_c^0$ states observed by the LHCb collaboration, namely the $\Omega_c^0(3000)$ ($\Gamma=4.5\pm0.9$), the  $\Omega_c^0(3050)$ ($\Gamma=0.8\pm0.3$) and the  $\Omega_c^0(3066)$  ($\Gamma=3.5\pm0.6$), although some fine-tuning of the model is still needed. By examining the corresponding couplings and compositions, we  
would conclude that the $\Omega_c^0(3000)$ resonance qualifies as a $\bar{K}\Xi_c$ resonance, the  
$\Omega_c^0(3050)$ is essentially a $D\Xi$ quasi-bound state and the $\Omega_c^0(3066)$ also has large $D\Xi$ components but strongly mixed with $\bar{K}\Xi'_c $ ones. We note that this third pole, located right below the $\bar{K}\Xi'_c $ threshold, is in fact a ``virtual state", as it appears in a Riemann sheet in which both $\bar{K}\Xi_c$ and $\bar{K}\Xi'_c $ loop functions are rotated, instead of appearing in the second Riemann sheet, which would imply the rotation of only  the $\bar{K}\Xi_c$ loop according to Eq.~(\ref{2ndRiemann}). This explains why the sum of compositeness to all meson-baryon states exceeds unity in this case. 

Far from being conclusive, these results demonstrate the need for a thorough investigation of all possible non-negligible terms of the meson-baryon interaction that may influence the generation of dynamical poles, paying a special attention to the dependence of the results on the assumed symmetries and on the free parameters of the theory \cite{albert}.

In the case of $VB$ scattering, ignoring the Born terms and employing subtraction constants mapped onto a cut-off of $\Lambda=800$\,MeV, we obtain spin degenerate $J^P=1/2^-,3/2^-$ resonances that follow a similar pattern as that found for the $PB$ case. A lower energy resonance mainly classifying as a $D^*\Xi$ molecule appears at 3231\,MeV and a higher energy resonance is generated at 3419\,MeV and corresponds to a $\bar{K}^*\Xi'_c $ composite state with some admixture of $\omega \Omega_c^0$ and $\phi \Omega_c$ components.  There is an additional pole in between these two, coupling strongly to $\bar{K}^*\Xi_c$ states.
These resonances cannot yet be identified with any reported state by the LHCb collaboration \cite{Aaij:2017nav}, as no significant peaks are seen standing out of the statistical noise in this energy region. 

We finally comment on the results in the bottom sector, obtained by replacing the charm mesons and baryons by their bottom counterparts in the meson--baryon interaction kernels obtained from the Lagrangians of Eqs.~(\ref{eq:vertexVPP})--(\ref{eq:vertexVVV}). We employ a factor $\kappa_b = 0.1$ in certain non-diagonal transitions that accounts for the much larger mass of the exchanged bottom vector mesons with respect to the light ones, analogously to $\kappa_c$. We find two narrow resonances at $6418$ MeV and  $6519$ MeV \cite{Montana:2017kjw}, quite above in energy from the four $\Omega_b^0$ excited states reported recently by LHCb lying in the range 6300--6350 MeV \cite{Aaij:2020cex}.  As argued in \cite{Liang:2020dxr} in connection to the states obtained in Ref.\cite{Liang:2017ejq}, it is quite improbable that dynamical meson-baryon models may be able to explain the observed $\Omega_b^0$ states, although an exception could be the model of Ref.~\cite{Nieves:2019jhp}, where a $\Omega_b^0$ of mass 6360 MeV, coupling strongly to $\bar{K}\Xi_b$ and belonging to a sexted of excited bottom states with $J^P=1/2^-$, has been predicted. Although not statistically significant, some structures can be hinted from the experimental spectrum in the higher energy region that may lead to clearer peaks if more data are gathered in future experiments. These structures would be then easily interpreted as having a molecular origin if tehir properties are close to those of the states predicted in Refs.~\cite{Montana:2017kjw,Liang:2017ejq}

\subsection{The $(I,S,C)=(\frac{1}{2},0,2)$ sector: $\Xi_{cc}$}
\label{sec:xicc}
The  discovery of a doubly charmed baryon $\Xi_{cc}^{++}$ by the LHCb Collaboration \cite{Aaij:2018gfl}, with a mass of $3621$ MeV, has motivated studies to search for possible doubly-charmed molecular states as that of Ref.~\cite{osetxi}, wherein several interesting predictions for quasi-bound states in the $C = 2$, $S = 0$ and $I = 1/2$ sector are presented, obtained following the same approach as in \cite{Debastiani:2017ewu}.
In the present study we extend the aforementioned method of Ref.~\cite{Montana:2017kjw} to this $\Xi_{cc}$ sector, both for pseudoscalar and vector mesons,  considering this new LHCb discovery as the ground state of the $\Xi^{++}_{cc}$. The coefficients of the interaction are presented in Table~\ref{taulaPB}. Note that we have also included a term weighted by a factor $ \xi_{cc}=(m_{\rho}/m_{J/\Psi})^2 \approx 1/16$ in the diagonal terms to account for the exchange contribution of the doubly-charm $J/\Psi$ meson.

\begin{table}[h!]
\centering
\scalebox{0.95}{
\begin{tabular}{lcccccccc}
\hline
& $\pi\Xi_{cc}$ & $D\Lambda_c$ & $\eta\Xi_{cc}$ & $K\Omega_{cc}$ & $D\Sigma_c$ & $D_s\Xi_c$ & $D_s\Xi_c'$ & $\eta'\Xi_{cc}$\\
\hline
$\pi\Xi_{cc}$ & 2 & $\frac{3}{2}\kappa_c$ & 0 & $\sqrt{\frac{3}{2}}$ &  $\frac{-1}{2}\kappa_c$ & 0 & 0 & 0\\
$D\Lambda_c$ & & $1 - \xi_{cc}$ & $\frac{-1}{2}\kappa_c$ & 0 & 0 & 1 & 0 & $\frac{-1}{\sqrt{2}}\kappa_c$\\
$\eta\Xi_{cc}$ & & & 0 & $\sqrt{\frac{3}{2}}$ &  $\frac{-1}{2}\kappa_c$ & $\kappa_c$ & $\frac{1}{\sqrt{3}}\kappa_c$ & 0\\
$K\Omega_{cc}$ & & & & 1 & 0 & $\sqrt{\frac{3}{2}}\kappa_c$ & $\frac{-1}{\sqrt{2}}\kappa_c$ & 0\\
$D\Sigma_c$ & & & & & $3-\xi_{cc}$ & 0 & $\sqrt{3}$ & $\frac{-1}{\sqrt{2}}\kappa_c$\\
$D_s\Xi_c$ & & & & & & $1 - \xi_{cc}$ & 0 & $\frac{-1}{\sqrt{2}}\kappa_c$\\
$D_s\Xi_c'$ & & & & & & & $1 - \xi_{cc}$ & $\frac{-1}{\sqrt{6}}\kappa_c$\\
$\eta'\Xi_{cc}$ & & & & & & & & 0  \\
\hline
\end{tabular}
}
\caption{$C_{ij}$ coefficients for $PB$ scattering in the $(I,S,C)=(\frac{1}{2},0,2)$ sector}
\label{taulaPB}
\end{table}

\begin{figure}[h!]
\centering
\includegraphics[width=3.4 in]{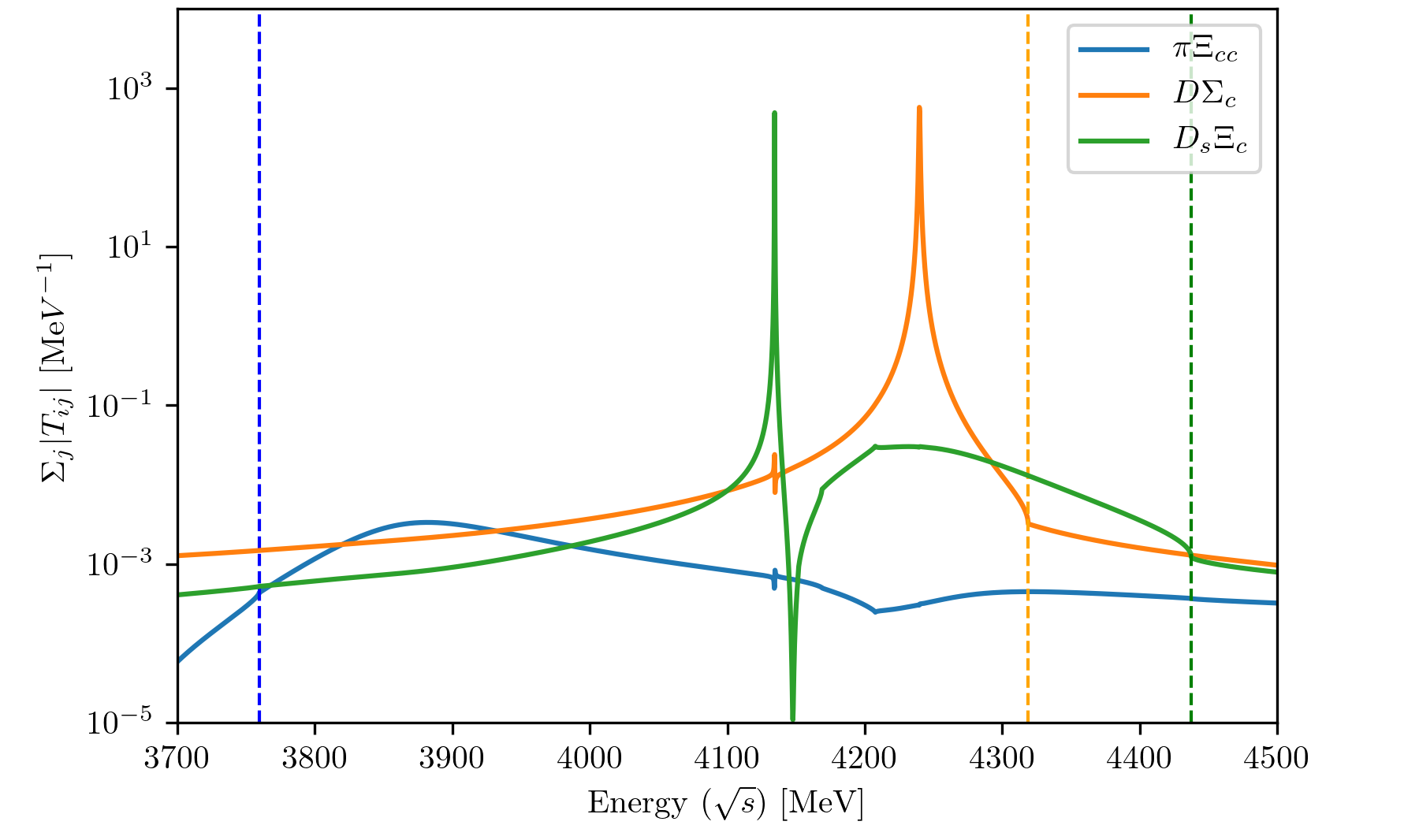}
\caption{The solid lines represent the sum over all $j$ channels of the module of the $PB$ scattering amplitude $T_{ij}$, for $\Lambda = 800$ MeV and $i = \pi\Xi_{cc},\  D\Sigma_c, \ D_s\Xi_c$. The dashed vertical lines correspond to the threshold masses of the respective channels.}
\label{grafica}
\end{figure}

Some indications about the location of possible poles in the complex plane may be found by examining the dependence of the scattering amplitude on the real axis. The quantity $\sum_j |T_{ij}|$ is shown in Fig.~\ref{grafica} for $i = \pi\Xi_{cc}, \ D\Sigma_c, \ D_s\Xi_c$. We observe the presence of two narrow states, coupling strongly to $D_s\Xi_c$ (green line) and  $D\Sigma_c$ (orange line) states. In addition, we observe two broad structures, one at lower energy, clearly visible in the amplitudes involving the $\pi\Xi_{cc}$ channel (blue line), and another at higher energy, coupling considerably to $D_s\Xi_c$. The explicit poles in the complex plane are displayed in Table~\ref{resPB800}. As expected, we first find a wide resonance at $3862$ MeV, with a large width of $158$ MeV, as its mass is well over the $\pi\Xi_{cc}$ channel threshold to which it couples most. Next, we find a narrow state at $4135$ MeV, right below the $D\Lambda_c$ threshold, to which it couples most. This state also couples significantly to $\eta\Xi_{cc}$, $K\Omega_{cc}$, and $D_s\Xi_c$, and this is the reason for its clear appearance in the amplitudes involving this later state in Fig.~\ref{grafica}. We find a third state at 4239 MeV, which mostly qualifies as a $D\Sigma_c$ bound state. It is very narrow because it couples negligibly to the meson-baryon states allowed for its decay. 
Finally, another wide resonance is seen at $4240$ MeV. It is mostly a $D_s\Xi_c$ quasi-bound states but it couples strongly to all lower-lying channels, hence its large width of $166$ MeV.

\begin{table}[h!]
\centering
\begin{tabular}{lccccc}
 
\hline 
& &   &    &   & \\   [-3mm]
\multicolumn{6}{c}{ {\bf $0^- \otimes 1/2^+$} interaction in the $(I,S,C)=(\frac{1}{2},0,2)$ sector} \\
& &   &    &   & \\   [-3mm]
\hline
& &   &    &   & \\   [-3mm]
$M\;\rm[MeV]$       &   & $\qquad 3861.9  $    &    $\qquad 4134.5 $    &  $\qquad 4239.4$ &  $\qquad 4240.4$   \\
$\Gamma\;\rm[MeV]$  &  &  $\qquad 158$    &    $\qquad  0$    &  $\qquad 0.76$ &  $\qquad 166$   \\ 
\hline
  & $a_l$  &   $\qquad | g_i|$    &   $\qquad | g_i|$   &   $\qquad | g_i|$    &  $\qquad | g_i|$  \\  
${\pi\Xi_{cc}}(3759)$ &-2.92 &\qquad {2.10} & \qquad 0.03 & \qquad 0.01 & \qquad 0.83 \\
${D\Lambda_c}(4152)$ & -2.21 & \qquad {1.29} & {\qquad 1.73} & \qquad 0.01 & \qquad {1.13} \\
${\eta\Xi_{cc}}(4169)$ &  -2.75 & \qquad 0.02 & \qquad 0.83 & \qquad 0.09 & \qquad 0.92 \\
${K\Omega_{cc}}(4208)$ & -2.81 & \qquad {1.22} & \qquad {1.10} & \qquad 0.12 & \qquad 0.78 \\
${D\Sigma_c}(4319)$ & -2.27  & \qquad 0.23 & \qquad 0.08 & \qquad {2.88} & \qquad 0.01 \\
${D_s\Xi_c}(4438)$ & -2.31 & \qquad 0.36 & \qquad {1.09} & \qquad 0.02 & \qquad {4.07} \\
${D_s\Xi_c'}(4545)$ & -2.34 & \qquad 0.18 & \qquad 0.01 & \qquad {1.55} & \qquad 0.01\\
${\eta'\Xi_{cc}}(4579)$ & -2.67 &  \qquad 0.03 & \qquad 0.20 & \qquad 0.11 & \qquad 0.33 \\
\hline
\end{tabular}
\caption{Position ($\sqrt{s}=M-\I\Gamma/2$) and couplings of the $PB$ $\Xi_{cc}$ states ($J^P = 1/2^-$)}
\label{resPB800}
\end{table}

\begin{figure}[b!]
\centering
\includegraphics[width=3.3 in]{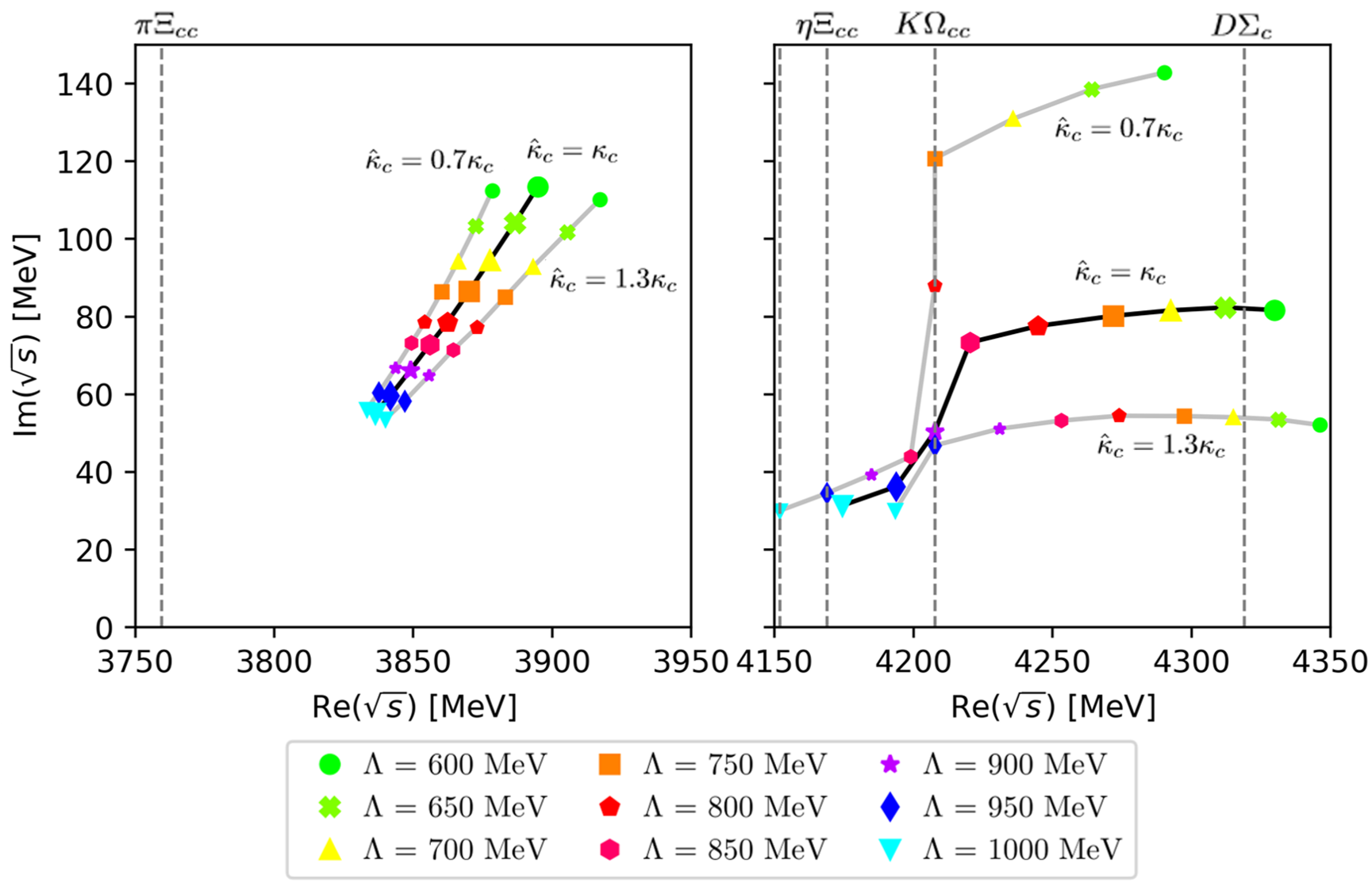}
\caption{The solid black lines represent the evolution of the poles positions as a function of the cut-off $\Lambda$. The solid gray lines represent the same evolution but considering an additional $SU(4)$ symmetry breaking of $\pm 30\%$. The dashed vertical lines correspond to the threshold masses of the stated channels. Left panel: resonance mainly coupled to $\pi\Xi_{cc}$. Right panel: resonance mainly coupled to $D_s\Xi_c$.}
\label{Analisi}
\end{figure}

Similarly to our study of the $\Omega_c^0$ states, we have also investigated the 
dependence of the $\Xi_{cc}$ states on the cut-off $\Lambda$ employed to calculate the subtraction constants $a_l(\mu)$ as well as
on variations of the $SU(4)$ symmetry breaking factors, $\kappa_c$ and $\xi_{cc}$. The
solid black lines in Fig.~\ref{Analisi} show the evolution of the lower (left panel) and higher energy poles (right panel) as the value of the cut-off is varied from $\Lambda = 600$ MeV to $\Lambda = 1000$ MeV in steps of $50$ MeV. The upper and lower solid gray lines represent the same evolution but reducing or increasing by 30\%, respectively, the values of $\kappa_c$ and $\xi_{cc}$ that modulate the SU(4) breaking in the kernel coefficients. We observe that when the cut-off $\Lambda$ increases, the resonances appear at lower energy. This is due to the consideration of an increased model space in the unitarization procedure, hence generating ``bound" states with larger binding energies. Moreover, when a resonance shifts to lower energies, it also becomes narrower in general. The steady decrease in width of the lower energy resonance (left panel) just reflects the loss of phase space of  $\pi\Xi_{cc}$ states, which are the only ones it can decay to. The behavior of the higher energy resonance (right panel) with increasing cut-off is qualitatively very different, showing a sudden drop in width around the $K\Omega_{cc}$ threshold mass as this channel is no longer reachable for decay. The evolution of the two other narrow $\Xi_{cc}$ resonances found in the present model is not represented in Fig.\ref{Analisi}, as it is essentially featureless. The resonances lower their masses as $\Lambda$ increases, but keep having a very small width as their couplings to the opened channels are essentially negligible. The dependence on the variation of the SU(4) parameters is much stronger in the higher energy resonance. The reason lies in the fact that this is a state strongly coupled to all the lower channels available for decay. The decrease (increase) of the SU(4) parameters makes these couplings stronger (weaker), therefore producing sensibly wider (narrower) resonances.

\begin{figure}[b!]
\centering
\includegraphics[width=3.2 in]{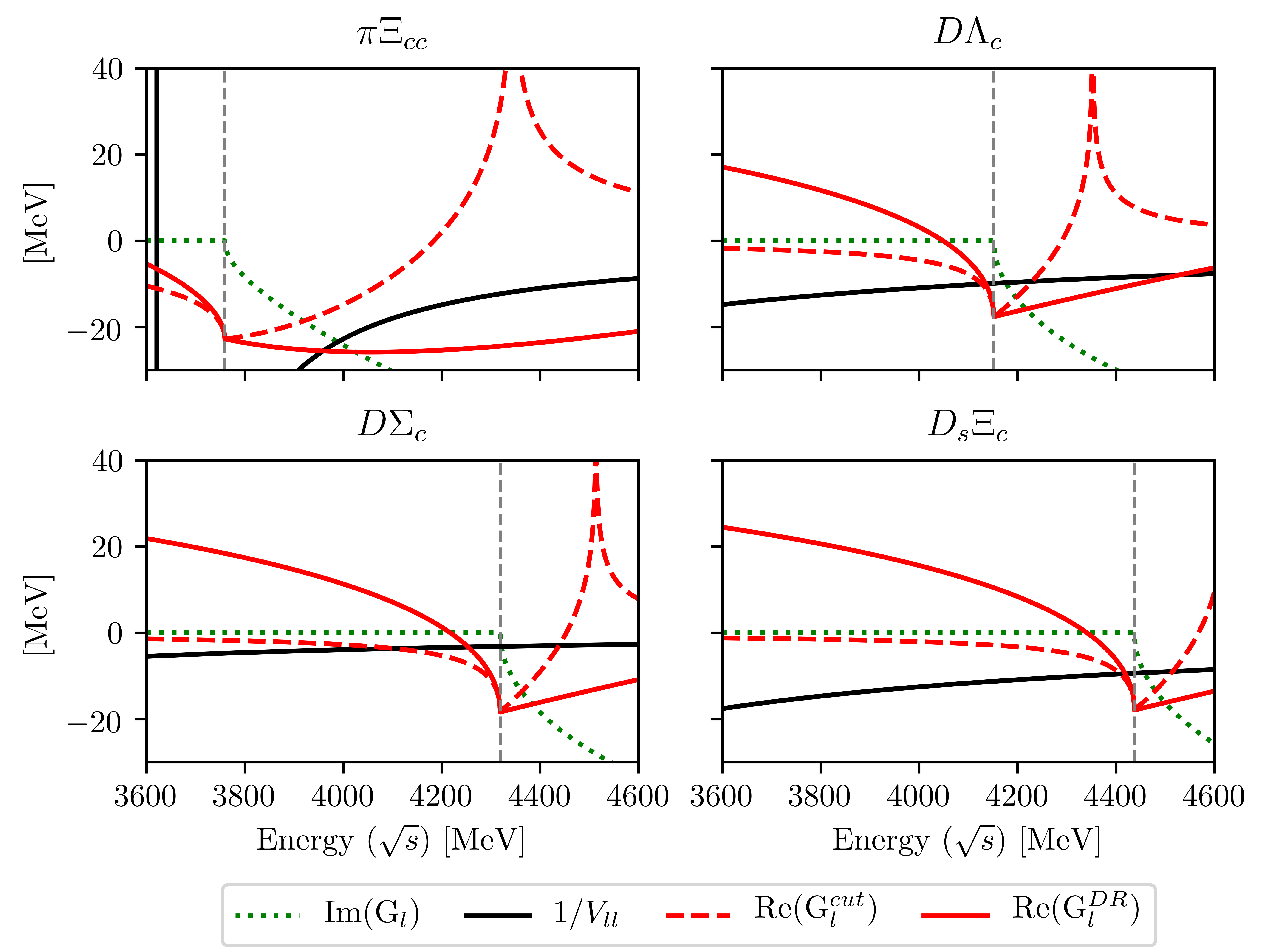}
\caption{The solid black line represents the inverse of the diagonal term of the channel $l$ potential, $1/V_{ll}$. As for the loop function $G_l$, the dotted green line represents its imaginary part, the solid red line its real part calculated via dimensional regularization with $\Lambda = 650$ MeV, and the dashed red line its real part calculated with a cut-off of 650 MeV. The dashed vertical lines correspond to the threshold masses of the respective channels.}
\label{CutOffDR}
\end{figure}

Comparing with the $PB$ scattering results of \cite{osetxi}, we observe a qualitative agreement on the nature and couplings of the states found, though not quantitative. In general, lower masses are obtained in \cite{osetxi}, by a few tens of MeV in the case of the
$\pi\Xi_{cc}$ and $D\Lambda_c$ states but by 150 MeV in the case of the $D\Sigma_c$ state.
In addition, the state coupling strongly to $D_s\Xi_c$ is missing from the results of \cite{osetxi}.
Differences in the $C_{ij}$ coefficients and in the $SU(4)$ symmetry breaking considered by both models have little effect on the final results, since diagonal couplings are identical and the non-diagonal ones have a weaker influence on the position of the poles.
Instead, we have seen that the main origin of the discrepancies observed in the three matching states found on both studies lies on the different regularization method employed. Requiring the loop integral to be regularized by a cut-off has lead us to obtain very similar results to those in \cite{osetxi}. This is better illustrated by examining Fig. \ref{CutOffDR}, where one sees that the real part of the loop function $G_l$ can have a different behaviour depending on the regularization scheme used. The real part of the cut-off loop functions below threshold are kept below zero, as opposed to their dimensional regularization counterparts. Moreover, above threshold, the cut-off loop functions present a cusp originating from the abrupt suppression of intermediate momentum states above $\Lambda$. These differences can have implications on the position of the poles since, in the simplified case with uncoupled channels and real energies, one would find the poles at the energies below threshold fulfilling $1/V_{ll} = G_l$. Notice that the intersection points between the inverse of the potential and the two different regularized loop functions are very close in the case of $D\Lambda_c$, while quite far in $D\Sigma_c$, explaining the differences found for these two resonances between both studies. This reasoning cannot be applied to the $\pi\Xi_{cc}$ resonance, since it is found above its threshold energy, but it should be valid for the $D_s\Xi_c$ pole. However, the $D_s\Xi_c$ resonance was not found in \cite{osetxi}.
Again, since the channels are coupled, one should be careful when drawing conclusions out of studying the uncoupled case, but it seems clear that the regularization scheme employed can have a strong impact on the position of the molecular states found. 

As for the results concerning the $VB$ scattering, giving rise to $J^P = 1/2^-, 3/2^-$ excited molecular states, we also find four poles. 
Although, in general, these poles present strong similarities with their $PB$ counterparts, there are noticeable differences probably due to the different order in energy on which the isospin equivalent meson-baryon states appear. As seen in Table~\ref{resVB800}, the first two resonances seem to have interchanged their role when comparing with the $PB$ states of Table~\ref{resPB800}, as the first one, at 4280 MeV, couples now more strongly to $D^*\Lambda_c$ states and the second one, at 4353 MeV, couples more strongly to $\rho\Xi_{cc}$.
The third pole is a narrow resonance strongly coupled to the $D^*\Sigma_c$ channel, and also mildly to $D^*_s\Xi_c'$. Both thresholds are higher than the resonance mass, which is approximately $4387$ MeV, hence its small width of only $1.3$ MeV. Comparing to the $D\Sigma_c$ $PB$ resonance, it seems that they could be spin partners of the same molecule.
Finally, we obtain a wide resonance at $4523$ MeV, coupling mostly to $K^*\Omega_{cc}$, but also to $D^*_s\Xi_c$, $\phi\Xi_{cc}$ and $\omega\Xi_{cc}$, the later channel being open for decay. 

A similar comparison with \cite{osetxi} holds for the $VB$ molecules. Ther is in general a qualitative agreement, but a quantitative disagreement on the values of the masses and widths. Again, only three $VB$ states were found in \cite{osetxi}, with comparable couplings to those of our states.

\begin{table}[h!]
\centering
\begin{tabular}{lccccc}
 
\hline 
& &   &    &   & \\   [-3mm]
\multicolumn{6}{c}{ {\bf $1^- \otimes 1/2^+$} interaction in the $(I,S,C)=(\frac{1}{2},0,2)$ sector} \\
& &   &    &   & \\   [-3mm]
\hline
& &   &    &   & \\   [-3mm]
\hline
$M\;\rm[MeV]$       &   & $\qquad 4279.8$ & $\qquad {4352.6}$ & $\qquad {4386.8}$ & $\qquad {4522.6}$ \\
$\Gamma\;\rm[MeV]$  &  & $\qquad{6.4}$ & $\qquad {22}$ & $\qquad {1.30}$ & $\qquad{54} $ \\
\hline
  & $a_l$  &   $\qquad | g_i|$    &   $\qquad | g_i|$   &   $\qquad | g_i|$    &  $\qquad | g_i|$  \\  
${D^*\Lambda_c}(4293)$ &  -2.25 & \qquad {1.67} & \qquad 0.86 & \qquad 0.08 & \qquad 0.21\\
${\rho\Xi_{cc}}(4393)$ &  -2.70  & \qquad {1.57} & \qquad {1.91} & \qquad 0.21 & \qquad 0.36\\
${\omega\Xi_{cc}}(4404)$ & -2.70  & \qquad 0.30 & \qquad 0.15 & \qquad 0.19 & \qquad 0.92\\
${D^*\Sigma_c}(4460)$ & -2.31  & \qquad 0.15 & \qquad 0.62 & \qquad {3.12} & \qquad 0.02\\
${D^*_s\Xi_c}(4581)$ & -2.34  & \qquad {1.45} & \qquad {1.76} & \qquad 0.23 & \qquad {1.64}\\
${K^*\Omega_{cc}}(4606)$ & -2.72  & \qquad 0.37 & \qquad {1.54} & \qquad 0.34 & \qquad {2.02}\\
${\phi\Xi_{cc}}(4641)$ & -2.67  & \qquad 0.43 & \qquad 0.21 & \qquad 0.27 & \qquad {1.30}\\
${D^*_s\Xi_c'}(4689)$ & -2.39 & \qquad 0.04 & \qquad 0.16 & \qquad {1.89} & \qquad 0.20\\
\hline
\end{tabular}
\caption{Position ($\sqrt{s}=M-\I\Gamma/2$) and couplings of the $VB$ $\Xi_{cc}$ states ($J^P = 1/2^-, 3/2^-$).}
\label{resVB800}
\end{table}





\section{Conclusions}

In this work we have focussed on the possible interpretation of some baryons, traditionally assigned to be 3-quark cluster structures, as being quasi-bound states (or ``molecules") emerging from a conveniently unitarized meson-baryon interaction in coupled channels. 

In particular, we have pointed out the importance of coupled channels to understand the differences between the charged and neutral final states in $K^0\Sigma$ photoproduction reactions off nucleons, measured at the ELSA facility. We have seen that the presence of a resonance predicted by our model at a mass of around 2~GeV can explain the sudden drop in the $\gamma p \to K^0\Sigma^+$ cross section and predict a peak in the neutral $\gamma n \to K^0\Sigma^0$ one around that energy. The success lies in the significant coupling strength of the resonance to the $K^*\Lambda$ and $K^*\Sigma$ channels, producing particular interference patterns that are able to accommodate  the observations, hence demonstrating the power of coupled-channel models and revealing the existence of a $N^*$ resonance around 2 GeV having the structure of a vector meson-baryon molecule.

We have extended the formalism to the description of baryons containing charm, motivated by the large amount of new states that are being discovered at higher energy facilities. In the 
$(I=0,S=-2,C=1)$ sector we find two resonances having energies and widths very similar to some of the $\Omega_c^0$ states discovered recently by LHCb, to which we would be assigning a spin-parity $J^P=1/2^-$. An experimental confirmation of this assignment would strongly enforce the interpretation of these states as meson-baryon molecules. In the bottom sector we obtain several $\Omega_b^-$ resonances in the energy region $6400$--$6800$\,MeV having a molecular meson--baryon structure. They do not correspond the recent states reported by LHCb at lower energies, but some hints of the predicted states could lie in the non-statistically significant structures seen at higher energies.

Finally, we also present predictions of possible molecular states having double charm content. We focus on the $(I=1/2,S=-1,C=2)$ sector and find possible molecular states with  $J^P=1/2^-,3/2^-$, which would be excited states of the ground state $\Xi_{cc}^{++}$ baryon recently identified by the LHCb Collaboration. The comparison of our results with experiment is expected to be done soon, as new analyses at higher energies are underway.

Although further tests and improvements are needed, it is clear that the interpretation of many baryons as being molecular states, hence departing from the conventional $qqq$ picture, is a solid prediction of the unitarized meson-baryon interaction models. In order to disentangle the dominant type of component, efforts should also be focussed on the study of decay rates, the analyses of spectral shapes in reactions involving these states, and the search of multiplet partners. The availability of data in the charm and bottom sectors has injected a renewed interest in the field, as it has provided unambiguous proof that some baryons, such as the $P_c(4312)^+$ and  $P_c(4450)^+$, receive a more natural interpretation as pentaquark systems. The internal structure of these and other possibly exotic baryons is yet to be clarified and, in this respect, valuable information is expected from the results of lattice QCD simulations that are becoming more sophisticated and precise in the last years.




\end{document}